\documentclass[pre,showpacs]{revtex4}

\usepackage{amsmath}
\usepackage{graphicx}

\usepackage[breaklinks]{hyperref}

\newcommand{\wraprefprepost}[3]{\wrapprepost{#1}{#2}{\ref{#3}}}
\newcommand{\formatrefplain}{\ref}
\newcommand{\formatrefparens}{\wraprefprepost{(}{)}}

\newcommand{\wrapprepost}[3]{{#1}{#3}{#2}}

\newcommand{\tagwithlabel}[2]{#1~#2}

\newcommand{\makelabeledcrossrefmacro}[4]
	{\newcommand{#3}{#1{#4}{#2}}}
\newcommand{\makecrossrefmaker}[3]
	{\newcommand{#1}{\makelabeledcrossrefmacro{#2}{#3}}}

\newcommand{\eqnrefformat}{\formatrefparens}
\newcommand{\eqnlabelbinding}{\tagwithlabel}
\newcommand{\eqnlabel}{eq.}
\newcommand{\Eqnlabel}{Eq.}
\newcommand{\eqnslabel}{eqs.}
\newcommand{\Eqnslabel}{Eqs.}

\newcommand{\eqnnum}{\eqnrefformat}
\makecrossrefmaker{\newlabeledeqnref}{\eqnlabelbinding}{\eqnnum}
\makecrossrefmaker{\newwordpluseqnref}{\tagwithlabel}{\eqnnum}

\newlabeledeqnref{\eqn}{\eqnlabel}
\newlabeledeqnref{\Eqn}{\Eqnlabel}
\newlabeledeqnref{\eqns}{\eqnslabel}
\newlabeledeqnref{\Eqns}{\Eqnslabel}

\newwordpluseqnref{\andeqn}{and}
\newwordpluseqnref{\througheqn}{through}

\newcommand{\figrefformat}{\formatrefplain}
\newcommand{\figlabelbinding}{\tagwithlabel}
\newcommand{\figlabel}{fig.}
\newcommand{\Figlabel}{Fig.}
\newcommand{\figslabel}{figs.}
\newcommand{\Figslabel}{Figs.}

\newcommand{\fignum}{\figrefformat}
\makecrossrefmaker{\newlabeledfigref}{\figlabelbinding}{\fignum}
\makecrossrefmaker{\newwordplusfigref}{\tagwithlabel}{\fignum}

\newlabeledfigref{\fig}{\figlabel}
\newlabeledfigref{\Fig}{\Figlabel}
\newlabeledfigref{\figs}{\figslabel}
\newlabeledfigref{\Figs}{\Figslabel}

\newwordplusfigref{\andfig}{and}
\newwordplusfigref{\throughfig}{through}

\newcommand{\sxnrefformat}{\formatrefplain}
\newcommand{\sxnlabelbinding}{\tagwithlabel}
\newcommand{\sxnlabel}{section}
\newcommand{\Sxnlabel}{Section}
\newcommand{\sxnslabel}{sections}
\newcommand{\Sxnslabel}{Sections}

\newcommand{\sxnnum}{\sxnrefformat}
\makecrossrefmaker{\newlabeledsxnref}{\sxnlabelbinding}{\sxnnum}
\makecrossrefmaker{\newwordplussxnref}{\tagwithlabel}{\sxnnum}

\newlabeledsxnref{\sxn}{\sxnlabel}
\newlabeledsxnref{\Sxn}{\Sxnlabel}
\newlabeledsxnref{\sxns}{\sxnslabel}
\newlabeledsxnref{\Sxns}{\Sxnslabel}
	
\newwordplussxnref{\andsxn}{and}
\newwordplussxnref{\throughsxn}{through}

\newcommand{\tblrefformat}{\formatrefplain}
\newcommand{\tbllabelbinding}{\tagwithlabel}
\newcommand{\tbllabel}{table}
\newcommand{\Tbllabel}{Table}
\newcommand{\tblslabel}{tables}
\newcommand{\Tblslabel}{Tables}

\newcommand{\tblnum}{\tblrefformat}
\makecrossrefmaker{\newlabeledtblref}{\tbllabelbinding}{\tblnum}
\makecrossrefmaker{\newwordplustblref}{\tagwithlabel}{\tblnum}

\newlabeledtblref{\tbl}{\tbllabel}
\newlabeledtblref{\Tbl}{\Tbllabel}
\newlabeledtblref{\tbls}{\tblslabel}
\newlabeledtblref{\Tbls}{\Tblslabel}
	
\newwordplustblref{\andtbl}{and}
\newwordplustblref{\throughtbl}{through}

\usepackage{xspace}

\newcommand{\apriori}{\textit{a priori}\xspace}
\newcommand{\ie}{i.e.}
 
\newcommand{\eg}{e.g.}

\newcommand{\GAH}{GAH\xspace}
\newcommand{\LPA}{LPA\xspace}
\newcommand{\LPAr}{LPAr\xspace}
\newcommand{\LPAm}{LPAm\xspace}
\newcommand{\LPAb}{LPAb\xspace}

\newcommand{\vertex}[1]{vertex~\( #1 \)}

\newcommand{\largeparens}[1]{\left ( #1 \right )}

\newcommand{\largecurly}[1]{\left \{ #1 \right \}}
\newcommand{\largeangle}[1]{\left \langle #1 \right \rangle }
\newcommand{\largestraight}[1]{\left | #1 \right |}

\newcommand{\half}{\frac{1}{2}}
\newcommand{\krondelta}[2]{\ensuremath{\delta\largeparens{#1,#2}}} 
\newcommand{\set}{\largecurly}

\newcommand{\argmax}[1]{\mathop{\mathrm{argmax}}_{#1}}
\newcommand{\mean}[1]{\largeangle{#1}}

\newcommand{\intersection}{\cap}

\newcommand{\mathpuncspace}{\quad}
\newcommand{\mathcomma}{\mathpuncspace,}
\newcommand{\mathperiod}{\mathpuncspace.}

\newcommand{\mat}[1]{\mathbf{#1}}

\newcommand{\numvertices}{\ensuremath{n}}
\newcommand{\numedges}{\ensuremath{m}}
\newcommand{\degree}[1]{\ensuremath{k_{#1}}}

\newcommand{\numredvertices}{\ensuremath{p}}
\newcommand{\numbluevertices}{\ensuremath{q}}
\newcommand{\reddegree}{\degree}
\newcommand{\bluedegree}[1]{\ensuremath{d_{#1}}}

\newcommand{\adjmat}{\mat{A}}
\newcommand{\adjelem}[2]{A_{#1#2}}

\newcommand{\probelem}[2]{P_{#1#2}}

\newcommand{\modularity}{\ensuremath{Q}}
\newcommand{\modularitynorm}{\frac{1}{2\numedges}}

\newcommand{\bpmodularity}{\ensuremath{Q^{\mathcal{B}}}}
\newcommand{\bpmodularitynorm}{\frac{1}{\numedges}}

\newcommand{\module}[1]{\ensuremath{g_{#1}}} 

\newcommand{\samemodule}[2]{\krondelta{\module{#1}}{\module{#2}}}

\newcommand{\vertlabel}[1]{\ensuremath{l_{#1}}}
\newcommand{\newvertlabel}[1]{\ensuremath{l^{\prime}_{#1}}}
\newcommand{\neighbors}[1]{\ensuremath{\sigma\largeparens{#1}}}
\newcommand{\objfunc}{\ensuremath{H}}
\newcommand{\constrainedobjfunc}{\ensuremath{H^\prime}}
\newcommand{\penaltyterm}{\ensuremath{G}}
\newcommand{\fapenaltyterm}{\ensuremath{G_1}}
\newcommand{\modpenaltyterm}{\ensuremath{G_2}}
\newcommand{\bppenaltyterm}{\ensuremath{G_3}}
\newcommand{\penaltywt}{\ensuremath{\lambda}}

\newcommand{\vertsum}[1]{\sum_{#1=1}^{\numvertices}}
\newcommand{\vertnbrsum}[2]{\sum_{#1 \in \neighbors{#2}}}
\newcommand{\excvertsum}[2]{\sum_{#1 \neq #2}}
\newcommand{\samelabel}[2]{\krondelta{\vertlabel{#1}}{\vertlabel{#2}}}

\newcommand{\arbsymelem}[2]{B_{#1 #2}}
\newcommand{\arbconst}{\ensuremath{C}}
\newcommand{\labeldegree}[1]{\ensuremath{K_{\vertlabel{#1}}}}
\newcommand{\numnbrswithlabel}[2]{N_{#1 \vertlabel{#2}}}

\newcommand{\order}[1]{\ensuremath{O\largeparens{#1}}}

\newcommand{\redlabeldegree}{\labeldegree}
\newcommand{\bluelabeldegree}[1]{\ensuremath{D_{\vertlabel{#1}}}}

\newcommand{\problinkintra}{p_{\mathrm{in}}}
\newcommand{\problinkinter}{p_{\mathrm{out}}}
\newcommand{\meandegree}{\mean{\degree{}}}
\newcommand{\meanoutdeg}{z_{\mathrm{out}}}

\newcommand{\schemex}{\ensuremath{X}}
\newcommand{\schemey}{\ensuremath{Y}}
\newcommand{\commsetx}{\ensuremath{C_x}}
\newcommand{\commsety}{\ensuremath{C_y}}
\newcommand{\probability}[1]{\ensuremath{P\largeparens{#1}}}
\newcommand{\mutinfosymbol}{\ensuremath{I}}
\newcommand{\mutinfo}[2]{\mutinfosymbol\largeparens{#1, #2}}
\newcommand{\normalizedmutinfosymbol}{\ensuremath{I_\mathrm{norm}}}
\newcommand{\normalizedmutinfo}[2]{\normalizedmutinfosymbol\largeparens{#1, #2}}
\newcommand{\entropysymbol}{\ensuremath{H}}
\newcommand{\entropy}[1]{\entropysymbol\largeparens{#1}}
\newcommand{\relentsum}[3]{\sum_{#3} #1 \log \frac{#1}{#2}}
\newcommand{\entropysum}[2]{-\sum_{#2} #1 \log #1}

\newcommand{\normmutinfo}{\normalizedmutinfosymbol}

\renewcommand{\eqnlabel}{Eq.}
\renewcommand{\eqnslabel}{Eqs.}
\renewcommand{\Eqnlabel}{Equation}
\renewcommand{\Eqnslabel}{Equations}
\renewcommand{\figlabel}{Fig.}
\renewcommand{\figslabel}{Figs.}
\renewcommand{\Figlabel}{Figure}
\renewcommand{\Figslabel}{Figures}
\renewcommand{\sxnlabel}{section}
\renewcommand{\Sxnlabel}{Section}
\renewcommand{\sxnslabel}{sections}
\renewcommand{\Sxnslabel}{Sections}
\renewcommand{\tbllabel}{Table}
\renewcommand{\tblslabel}{Tables}

\graphicspath{{pdffigs/}{epsfigs/}{mpsfigs/}}

\begin{document}

\title{Detecting network communities by propagating labels under constraints}

\author{Michael J. Barber}
\affiliation{AIT Austrian Institute of Technology GmbH, Foresight \& 
Policy Development Department, Vienna, Austria}
\email{michael.barber@ait.ac.at}

\author{John W. Clark}
\affiliation{Department of Physics, Washington University, Saint Louis, MO}

\date{\today}

\begin{abstract}

We investigate the recently proposed label-propagation algorithm (\LPA)
for identifying network communities. We reformulate the \LPA as an
equivalent optimization problem, giving an objective function whose
maxima correspond to community solutions. By considering properties
of the objective function, we identify conceptual and practical
drawbacks of the label propagation approach, most importantly the
disparity between increasing the value of the objective function
and improving the quality of communities found. To address the
drawbacks, we modify the objective function in the optimization
problem, producing  a variety of algorithms that propagate labels
subject to constraints; of particular interest is a variant that
maximizes the modularity measure of community quality. Performance
properties and implementation details of the proposed algorithms
are discussed.
Bipartite as well as unipartite networks are considered.

\end{abstract}


\maketitle

\section{Introduction} \label{sec:intro}

There is great current interest in identifying communities in
networks. Informally, communities in networks, or graphs, are
subgraphs whose vertices are more strongly connected to one another
than to the vertices outside the subgraph. A variety of approaches
have been taken to make concrete the idea of communities, giving
rise to a number of efficient methods for community identification
(for useful overviews, see Refs.~\cite{New:2004b,DanDiaDucAre:2005,ForCas:2008}).

Recently, \citet{RagAlbKum:2007} have introduced a label-propagation
algorithm (\LPA) for identifying network communities.  Initially,
each vertex in the graph is assigned a unique numeric label. The
label for each vertex is replaced with the most frequent label from
its neighbors. Relabeling continues until a stable set of labels
is reached.  Network communities are defined as the sets of vertices
bearing the same labels.  The \LPA offers a number of desirable
qualities, including conceptual simplicity, ease of implementation,
and practical efficiency---the algorithm rapidly
\citep{RagAlbKum:2007,LeuHuiLioCro:2008} finds community assignments
of high quality, as measured by the popular modularity measure
\citep{NewGir:2004}.

The \LPA was originally presented operationally, with communities
defined as the outcome of a specific procedure.
In this work, we consider an equivalent mathematical formulation, 
in which community solutions are
understood in terms of optima of an objective function. 
We define an objective function \( \objfunc \) based on the number
of edges that connect vertices with identical labels, and show that 
the \LPA identifies local optima of \( \objfunc \).  This is formally equivalent to minimizing 
the Hamiltonian for a ferromagnetic Potts model  \cite{TibKer:2008}. 
The mathematical formulation exposes a number of interesting properties of 
the \LPA.  A feature of conspicuous importance is that the globally optimal solution for 
any network is the uninteresting trivial solution in which all vertices are
assigned the same label, with other solutions found by label propagation corresponding to
suboptimal local maxima of \( \objfunc \).

The objective function optimized by label propagation thus corresponds
poorly to our conceptual understanding of communities---an increase
in \( \objfunc \) need not produce what we would consider to be
better communities. In particular, attempts to improve on the label
propagation algorithm by facilitating its escape from local
maxima in \( \objfunc \) may be counterproductive. We demonstrate
that this can create practical difficulties for improvement upon the
standard \LPA.

We next consider adding a term to the original objective function
that penalizes undesirable solutions, producing  algorithms that
propagate labels subject to constraints. We examine several
possibilities for the penalty term.  Of special interest is a
penalty term that works to divide vertices into groups of equal
total degree, yielding a label propagation variant that strictly
maximizes the modularity \citep{NewGir:2004} while maintaining the
favorable computational complexity of the standard \LPA. We characterize 
the effectiveness of the several label propagation algorithms through 
application to a model network and a selection of real-world networks.

The structure of the remainder of the paper is as follows. In
\sxn{sec:lpa} we briefly summarize the original operational
presentation of the label propagation algorithm \citep{RagAlbKum:2007}.
In \sxn{sec:objfunc}, we reformulate the label propagation algorithm
as a mathematical optimization problem, and in \sxn{sec:drawbacks}
consider  drawbacks of the \LPA thus revealed. We address the
drawbacks in \sxn{sec:constrainedlpa} by adding constraints to the
optimization problem, with attendant notes on implementation in the
appendices. Performance of several label propagation 
variants are compared in \sxn{sec:modmax}, for both unipartite 
and bipartite networks.   We conclude with a summary
and discussion in \sxn{sec:discussion}.

\section{The Label Propagation Algorithm} \label{sec:lpa}

The identification of communities in networks is a topic of great
recent interest. Formulation of the problem presents two main challenges.
First, the notion of community is imprecise, requiring a definition
to be provided for what constitutes a community. Second, community
solutions must also be practically realizable for networks of
interest. The interplay between these challenges allows a variety
of community definitions and community identification algorithms
suited to networks of different sizes, as measured by the number
of vertices \( \numvertices \) or edges \( \numedges \) in the
network.

A prominent formulation of the community-identification problem is
based on the modularity \( \modularity \) introduced by
\citet{NewGir:2004}. The quality of communities given by a partition
of the network vertices is assessed by comparing the number of edges
between vertices in the same community to the number expected from
a null model network. Formally, this is
\begin{equation}
	\modularity = \modularitynorm \sum_{i,j} 
			\largeparens{\adjelem{i}{j} - \probelem{i}{j}} 
			\samemodule{i}{j}
	\mathcomma
	\label{eq:defmodularitymodel}
\end{equation}
where the \( \adjelem{i}{j} \) are components of the adjacency
matrix for the network, and \( \module{i} \) is the community for
vertex~\( i \). The presence of the Kronecker delta term 
\( \samemodule{i}{j} \) restricts the sum to edges 
within communities. The probability of an edge existing between 
vertices~\( i \) and~\( j \) in the null model network is given 
by  \( \probelem{i}{j} \). The standard choice of null model 
takes the probability of an edge to be proportional to the
product of the degrees \( \degree{i} \) and \( \degree{j} \) of the 
vertices, giving
\begin{equation}
	\probelem{i}{j} = \frac{\degree{i}\degree{j}}{2\numedges}
	\mathperiod
	\label{eq:standardnullmodel}
\end{equation}
With this choice for \( \probelem{i}{j} \), the modularity becomes
\begin{equation}
	\modularity = \modularitynorm \sum_{i,j} 
			\largeparens{\adjelem{i}{j} - \frac{\degree{i}\degree{j}}{2\numedges}} 
			\samemodule{i}{j}
	\mathperiod
	\label{eq:defmodularity}
\end{equation}
Communities are then sought by finding partitions of the set of
vertices that have a high value for modularity. The global maximum
of \( \modularity \) is generally inaccessible, as the number of
possible partitions for a set grows too rapidly to be feasibly examined for
all but the smallest networks, although effective heuristics exist for
finding high modularity solutions. A seminal example is the 
greedy agglomerative hierarchical  algorithm \citep{New:2004a,ClaNewMoo:2004}, wherein 
pairs of communities are successively merged so 
as to cause the largest possible increase in \( \modularity \)
at each step.

Recently, \citet{RagAlbKum:2007} have introduced a label-propagation
algorithm (\LPA) for identifying network communities. In contrast
to the above modularity-based approach, communities are defined in
the \LPA as vertex partitions identified by a specific algorithm.
The algorithm is conceptually simple in its operation.  Initially,
each vertex in the graph is assigned a unique numeric label. The
label for each vertex is then replaced with the most frequent label
amongst its neighbors; when several labels are equally frequent,
the current label is kept if it is among the most frequent, while
otherwise a new label is chosen at random from the most frequent.
Vertices are repeatedly relabeled, with the algorithm terminating
when the label for each vertex is (one of) the most frequent of the
labels for the neighbors of the vertex. To avoid possible cycles
and ensure termination, \citet{RagAlbKum:2007} suggest updating the
vertex labels asynchronously and in random order. Network communities
are then associated with sets of vertices bearing the same labels.

The \LPA offers a number of desirable qualities. As described above,
it is conceptually simple, being readily understood and quickly
implemented. Communities found can be of high quality, as
assessed, \eg, by the modularity. The algorithm is
efficient in practice. Each relabeling iteration through the vertices
has a computational complexity linear in the number of edges in the
graph. The total number of iterations is not \apriori clear, but
relatively few iterations are needed to assign the final label to
most of the vertices (over 95\% of vertices in 5 iterations,
see Refs.~\citep{RagAlbKum:2007,LeuHuiLioCro:2008}).

Two related works are of particular note. First, \citet{TibKer:2008}
have identified the label propagation algorithm as formally equivalent
to minimizing the Hamiltonian for a kinetic Potts model, and used
this to argue that, at least in some networks, the identified
communities may be meaningless. Additionally, through empirical
investigation of two real-world networks, \citeauthor{TibKer:2008}
have shown that the number of distinct community solutions may be
very large---much larger than the number of network vertices. Taken
together, these observations highlight the need for further assessment
of the quality of communities found using label propagation.

Second, \citet{LeuHuiLioCro:2008} have examined the \LPA as a basis
for analyzing large networks, focusing on performance
characteristics and limitations. They suggest a number of extensions
and optimizations, resulting in a modified algorithm that is able
to find communities in a network with tens of millions of edges
in a few minutes using a desktop PC. This study thus suggests that 
label propagation has tremendous potential as an effective and efficient 
method for community identification.

\section{An Objective Function for Label Propagation} \label{sec:objfunc}

Thus far, the \LPA has been presented operationally---the community
solutions are defined as the outcome of a specific procedure.
Alternatively, an equivalent mathematical formulation, first recognized 
by \citet{TibKer:2008}, can be given, where community solutions are 
understood in terms of the results of applying an optimization procedure 
to an objective function. The
optimization procedure is the \LPA, while the objective function
remains to be specified. The mathematical reformulation thus requires
defining the objective function, which provides an alternate means
of understanding solutions found by the \LPA. 

To effect this reformulation, we first express the \LPA optimization procedure as
\begin{equation}
	\newvertlabel{v} = \argmax{\vertlabel{}} \vertnbrsum{u}{v} \samelabel{u}{} 
	\label{eq:lparule}
	\mathcomma
\end{equation}
where \( \vertlabel{u} \) is the current label for \vertex{u}, 
\(\newvertlabel{v} \) is the new label for \vertex{v}, \( \neighbors{v}\) 
is the set of vertices neighboring \( v \) in the network, and
\( \delta \) is the Kronecker delta. In the event that multiple
values would maximize the sum, the result of \( \argmax{\vertlabel{}} \) 
should be taken as for the procedural description of \LPA, \ie,
keep the current label if it would satisfy \eqn{eq:lparule}, otherwise
take a label at random that satisfies \eqn{eq:lparule}.

\Eqn{eq:lparule} can be written in terms of the adjacency 
matrix \( \adjmat \) for the network, giving
\begin{equation}
	\newvertlabel{v} = \argmax{\vertlabel{}} \vertsum{u} \adjelem{u}{v}\samelabel{u}{} 
	\label{eq:lparuleadj}
	\mathcomma
\end{equation}
where \( \numvertices \) is the number of vertices in the network.
Consistent with the \LPA, the adjacency matrix elements 
\( \adjelem{u}{v} \) are all elements of \( \set{0, 1} \). However, 
the discrete nature of the \( \adjelem{u}{v} \) is never made use of, 
so the form in \eqn{eq:lparuleadj} is equally applicable to weighted
networks.

Next, we introduce an objective function \( \objfunc \) that is
maximized by the optimization procedure. Intuitively, we can view
the \LPA as working to assign labels so as to increase the number
of edges that connect vertices with identical labels. Formally,
this number has the expression
\begin{equation}
	\objfunc = \half \vertsum{v} \vertnbrsum{u}{v} \samelabel{u}{v}
	\mathperiod
	\label{eq:objfunctionbase}
\end{equation}
\Eqn{eq:objfunctionbase} can be rewritten in terms of the network 
adjacency matrix, giving 
\begin{equation}
	\objfunc = \half \vertsum{v} \vertsum{u} \adjelem{u}{v} \samelabel{u}{v}
	\mathperiod
	\label{eq:objfunction}
\end{equation}
We note that maximizing \( \objfunc \) is equivalent to minimizing 
the Hamiltonian for a ferromagnetic Potts model; this connection has been 
previously recognized by \citet{TibKer:2008}. The use of a Potts model 
Hamiltonian in network partitioning has been explored in depth 
by \citet{ReiBor:2006}.

It remains to be verified that the optimization rule in \eqn{eq:lparule}
does in fact maximize the objective function in \eqn{eq:objfunction}.
Consider updating the label for some \vertex{x}. We rewrite
\eqn{eq:objfunction} to treat \vertex{x} separately, yielding
\begin{equation}
	\objfunc = \half \largeparens{
		\excvertsum{v}{x} \excvertsum{u}{x}  \adjelem{u}{v} \samelabel{u}{v}
		+ \vertsum{u} \adjelem{u}{x} \samelabel{u}{x}
		+ \vertsum{v} \adjelem{x}{v} \samelabel{x}{v}
		- \adjelem{x}{x}
	}
	\mathperiod
	\label{eq:asymobjfuncverifymax}
\end{equation}
Taking advantage of the symmetry of the adjacency matrix, we can simplify \eqn{eq:asymobjfuncverifymax}, giving
\begin{equation}
	\objfunc = \half \largeparens{
	\excvertsum{v}{x} \excvertsum{u}{x}  \adjelem{u}{v} \samelabel{u}{v}
	- \adjelem{x}{x}
	}
	+ \vertsum{u} \adjelem{u}{x} \samelabel{u}{x}
	\mathperiod
	\label{eq:objfuncverifymax}
\end{equation}
The final term on the right hand side of \eqn{eq:objfuncverifymax}
is exactly of the form maximized by the \LPA optimization rule as 
expressed in
\eqn{eq:lparuleadj}, while the other terms are independent of the label
on \vertex{x}. Thus, the objective function never decreases under
the action of the \LPA, ultimately reaching a local maximum or limit
cycle.

An important property of the label propagation algorithm is immediately
apparent from the form of \( \objfunc \). For any network, the \LPA
allows an uninteresting trivial solution in which all vertices are
assigned the same label \citep{RagAlbKum:2007}. From \( \objfunc\), 
we see that the trivial solution is in fact the globally optimal
solution. Other solutions found by label propagation correspond to
local maxima of \( \objfunc \).

As the \LPA optimization procedure in \eqn{eq:lparuleadj} produces
only local changes, the search for maxima in \( \objfunc \) is
prone to becoming trapped at a local optimum instead of the global
optimum. While normally a drawback of local search algorithms, this
characteristic is essential to the function of the \LPA: the trivial
optimal solution is avoided by the dynamics of the local search
algorithm, rather than through formal exclusion.

\section{Drawbacks of Label Propagation} \label{sec:drawbacks}

The label propagation algorithm as a search scheme thus depends on
a certain degree of ineffectuality.  A typical way to attempt 
improvement of a local search algorithm is to make it more able to escape from
local maxima in \( \objfunc \). Such improvements to the \LPA may
be quite counterproductive, as better solutions in terms of 
\(\objfunc \)---notably, the global maximum---may be quite useless
in practical terms. Despite this, label propagation in practice 
can produce communities that are of high quality in terms of, \eg,
modularity: the local maxima are frustrated equilibria, with localized
groups of well-connected vertices having the same label and with
comparatively few edges between the groups.

Generally, there is a poor correspondence between \( \objfunc \)
and our conceptual understanding of communities. Maximizing 
\(\objfunc \), be it by label propagation or another approach, need
not produce better communities. Regardless, using the \LPA works by
maximizing \( \objfunc \), raising the question of whether, and in
what sense, we are improving community quality. Operationally, it
is again unclear what it might mean to try improving the \LPA. Does
improving the search efficacy actually give better communities? How
do we prevent our optimizations from reaching the global maximum
of \( \objfunc \), or other uninteresting solutions with high values
of \( \objfunc \)?

To illustrate the difficulties involved, we consider a possible
optimization of the label propagation algorithm. When a vertex label
is to be updated, it is necessary to handle the case where multiple
labels are equally frequent for the neighboring vertices. In the
standard \LPA, these ties are broken by keeping the current label
for the vertex, if it is one of the most frequent, or otherwise by
selecting a label at random from the most frequent. In our optimized
version, we will always select a label at random from the most
frequent; in light of this additional randomization, we denote the
modified algorithm as \LPAr. The tie-breaking rule for the standard
\LPA corresponds to halting when a plateau in the \( \objfunc \)
space is reached, while \LPAr corresponds to allowing a random walk
on the plateau in search of better solutions.

\begin{figure}
	\includegraphics[width=\columnwidth]{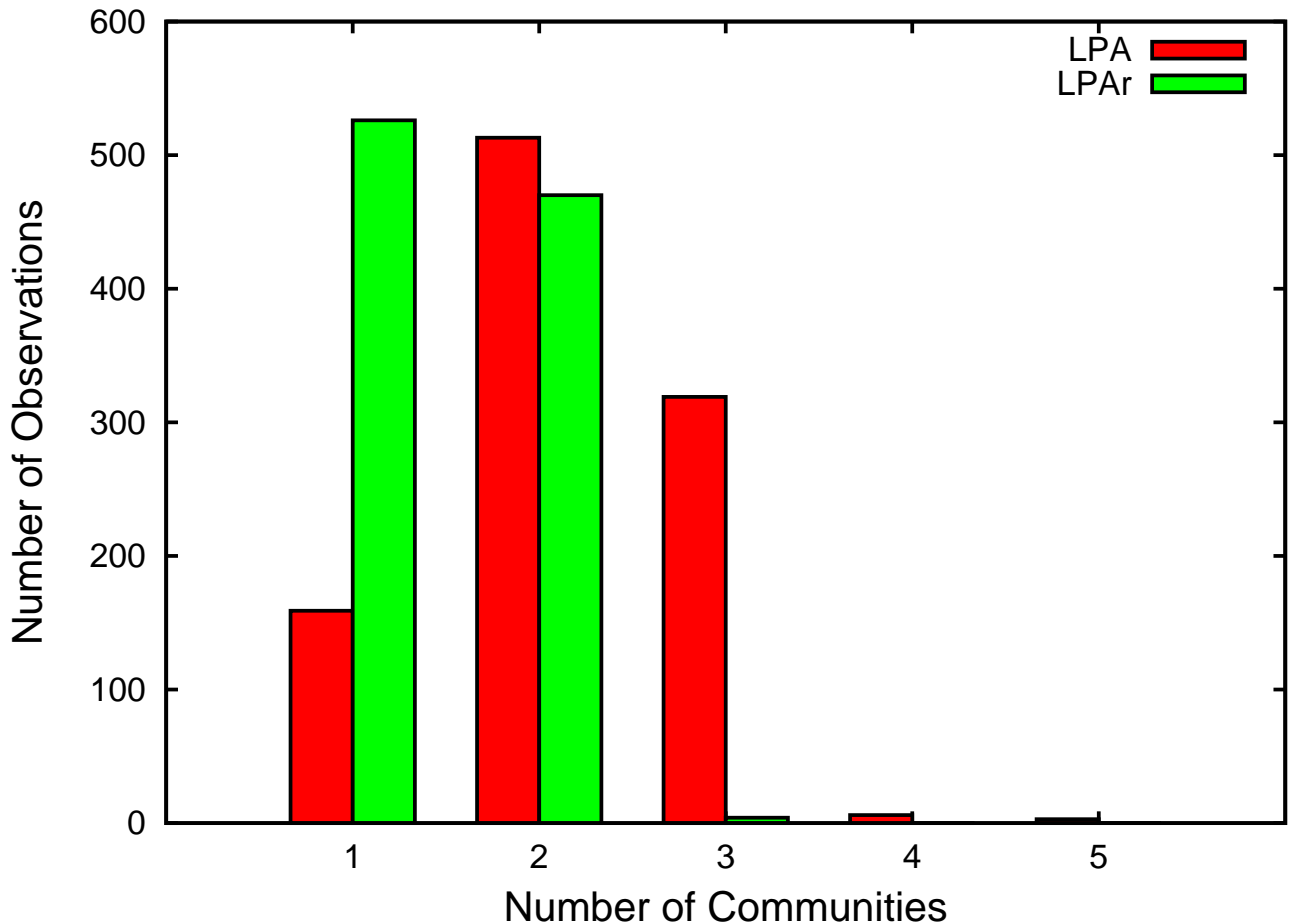}
	\caption{An attempted optimization of the label propagation algorithm
	produces dubious gains for the Southern women network. The modified
	\LPAr frequently produces the trivial solution, with all vertices
	assigned to the same community. In the network considered, we expect
	at least two communities based on the ethnographic study from which
	the data is drawn.  }
	\label{fig:optimizeLPA}
\end{figure}

In \fig{fig:optimizeLPA}, we show the number of communities found
for one thousand applications of the standard \LPA and the putatively
optimized \LPAr to networks derived from the Southern women
data. The data were collected by
\citet{DavGarGar:1941} as part of an extensive study of class and
race in the Deep South. The network represents interactions of a
group of 18 women at 14 various events in and around Natchez,
Mississippi during the 1930s. This much-studied network is typically
found to have two communities using methods of social network
analysis \citep{Fre:2003}, in accord with the conclusions from the
original ethnographic study. Unfortunately,
our attempted optimization has a perverse result with the Southern
women network: the principal effect of the optimization is to
drastically increase the frequency at which the algorithm assigns
the same label to all vertices, failing to capture any aspect of
the known community structure.

At least in the Southern women network, several practical drawbacks
arise from the key conceptual drawback discussed above. Optimization
is made difficult, as seen in this case based on comparison with a
known community structure. Further, the objective function optimized
by \LPA provides no mechanism for testing the quality of the resulting
community solutions---we must instead assess quality through 
auxiliary considerations such as the number of communities or, \eg, the
modularity \( \modularity \) of the community solution.

\section{Constrained Label Propagation} \label{sec:constrainedlpa}

A well-established approach for eliminating undesirable solutions
is to modify the objective function by adding a constraint term that penalizes
the undesirable solutions. Denoting the modified objective function
as \( \constrainedobjfunc \) and the penalty term as 
\( \penaltyterm\), we have
\begin{equation}
	\constrainedobjfunc = \objfunc - \penaltywt \penaltyterm
	\mathcomma
	\label{eq:penalizedobjfunc}
\end{equation}
where \( \penaltywt \) is a parameter that weights \( \penaltyterm \) against 
the original objective function \( \objfunc \).
Numerous choices are possible for \( \penaltyterm \); we consider
three possibilities below.

Within the specific area of communication identification, the
approach has been used at least since the landmark paper by
\citet{FuAnd:1986} applying methods of statistical mechanics to
combinatorial optimization problems, including
graph bipartitioning. We base a first penalty term \( \fapenaltyterm\) 
on their classic work. We seek to divide the vertices into groups
of the same size. In terms of the labels, we define
\begin{eqnarray}
	\fapenaltyterm & = & \half \vertsum{\vertlabel{}} \largeparens{ \vertsum{v} \samelabel{v}{} }^2 \nonumber \\
	& = & \half \vertsum{u} \vertsum{v} \samelabel{v}{u}
	\mathperiod
	\label{eq:fapenaltyterm}
\end{eqnarray}
The penalty term \( \fapenaltyterm \) produces the smallest value
when all vertices have unique labels, and the largest value when
all vertices have the same label. Thus, the trivial global optimum
of \( \objfunc \) is penalized and hopefully avoided.

Alternatively, following a strategy that mirrors contemporary methods for 
community detection, we can try to divide the vertices into groups which
have a similar total degree. We define a second penalty term 
\(\modpenaltyterm \) to capture this idea. The total degree \( \labeldegree{} \) of 
the vertices with a given label \( \vertlabel{} \) is
\begin{equation}
	\labeldegree{} = \vertsum{i} \degree{i} \samelabel{i}{}
	\mathcomma
	\label{eq:totaldegreeforlabel}
\end{equation}
where \( \degree{i} \) is the degree of \vertex{i}. 
A suitable definition for \( \modpenaltyterm \) is
\begin{equation}
	\modpenaltyterm = \half \vertsum{\vertlabel{}} \labeldegree{}^2
	\mathperiod
	\label{eq:modpenaltytermlabelvolume}
\end{equation}
As with 
\(\fapenaltyterm \), \( \modpenaltyterm \)  is minimal when all
vertices have unique labels and maximal when all vertices have the
same label, working to avoid the trivial global optimum.

We can rewrite \( \modpenaltyterm \) in the form
\begin{eqnarray}
	\modpenaltyterm & = & \half \vertsum{\vertlabel{}} \largeparens{ \vertsum{v} \degree{v} \samelabel{v}{} } ^ 2 \nonumber \\
	& = & \half \vertsum{u} \vertsum{v} \degree{u} \degree{v} \samelabel{u}{v}
	\mathperiod
	\label{eq:modpenaltyterm}
\end{eqnarray}
Incorporating \( \modpenaltyterm \) into \( \constrainedobjfunc \), 
we obtain
\begin{equation}
	\constrainedobjfunc = \half \vertsum{u} \vertsum{v} \largeparens{ \adjelem{u}{v} - \penaltywt  \degree{u} \degree{v} } \samelabel{u}{v}
	\mathperiod
	\label{eq:nearmodobjfunc}
\end{equation}
If we select
\begin{equation}
	\penaltywt = \frac{1}{2\numedges}
	\mathcomma
	\label{eq:modpenaltywt}
\end{equation}
where \( \numedges \) is the number of edges in the network, 
the objective function may be written as
\begin{equation}
	\constrainedobjfunc = \numedges \modularity
	\mathperiod
	\label{eq:modobjfunc}
\end{equation}
In \eqn{eq:modobjfunc}, \( \modularity \) is the standard modularity measure
\citep{NewGir:2004}. 

Recalling that the label propagation rule as
given by \eqn{eq:lparuleadj} requires only that a symmetric matrix
be used, we can see from \eqn{eq:nearmodobjfunc} that modularity
can be be locally maximized by the label propagation algorithm; we denote
this modularity-specialized algorithm as \LPAm. Implementation issues 
are described in appendix~\sxnnum{sec:lpamimplementation}. We note 
that \LPAm, due to the effect of \( \modpenaltyterm \), is well suited to 
aggressive optimization, but we do not pursue such optimizations in the 
present work.

The penalty term \( \modpenaltyterm \) plays the same role as the
null model network used to define the modularity 
(see, \eg, Ref.~\citep{NewGir:2004}). The idea holds quite generally: 
various null model networks could be used to define specialized 
modularity measures, or penalty terms could equivalently be 
introduced into the objective function. This allows the interesting historical
interpretation that \citet{FuAnd:1986} made use of a modularity
measure for community identification over two decades ago.

As a further example, we develop an analogous label propagation
algorithm to maximize a recently introduced \citep{Bar:2007} version
of modularity adapted to the important special class of bipartite
networks.  The vertices of a bipartite network can be partitioned
into two disjoint sets such that no two vertices within the same
set are adjacent; equivalently, the vertices in a bipartite graph
can be assigned one of two colors, say red and blue, with no edges
present between vertices bearing the same color.  There are thus
two distinct kinds of vertices, providing a natural representation
for many affiliation or interaction networks, with one kind of
vertex representing actors and the other representing relations.

The distinction between the two parts of the network can be
incorporated into a modularity measure by defining a suitable null
network model. In contrast to the standard choice given in
\eqn{eq:standardnullmodel}, the two kinds of vertices
must be treated separately, with non-zero probability of an edge only between 
vertices belonging to different parts of the network. For a red vertex \( i \) with 
degree \( \reddegree{i} \) and a blue vertex \( j \) with degree \( \bluedegree{j} \), 
the null model is defined so that 
\begin{equation}
	\probelem{i}{j} = \frac{\reddegree{i}\bluedegree{j}}{\numedges}
	\mathperiod
	\label{eq:bipartnullmodel}
\end{equation}
Using \eqn{eq:bipartnullmodel}, the bipartite modularity \( \bpmodularity \) 
is 
\begin{equation}
	\bpmodularity = \bpmodularitynorm \sum_{i,j}{}^{\prime} \largeparens{\adjelem{i}{j} -  \frac{\reddegree{i}\bluedegree{j}}{\numedges}} \samemodule{i}{j}
	\mathperiod
	\label{eq:bipartmoddefinition}
\end{equation}
The sums in \eqn{eq:bipartmoddefinition} are to be interpreted as running over the vertices in 
the two parts of the network, \ie, \( i \) is restricted to run over only the red vertices, 
while \( j \) is restricted to run over only the blue vertices.

For the present work, it is simpler to allow unrestricted sums over all the vertices. 
To do this, for each \vertex{v} we associate two degree measures, a red
degree \( \reddegree{v} \) and a blue degree \( \bluedegree{v} \).
If \vertex{v} is red, we require \( \bluedegree{v} = 0\), while if
it is blue we require \( \reddegree{v} = 0 \).  In either case, the
non-zero degree is the number of edges incident on the vertex. With
this construction, \eqn{eq:bipartmoddefinition} becomes
\begin{equation}
	\bpmodularity = \modularitynorm \vertsum{i} \vertsum{j} \largeparens{\adjelem{i}{j} - \frac{2\reddegree{i}\bluedegree{j}}{\numedges} } \samemodule{i}{j}
	\mathcomma
	\label{eq:bipartmodfullsums}
\end{equation}
where now the sums run over all vertices.

We now define a penalty term \( \bppenaltyterm\) 
for bipartite networks as
\begin{equation}
	\bppenaltyterm  = \half \vertsum{\vertlabel{}} \redlabeldegree{}\bluelabeldegree{}
	\mathcomma
	\label{eq:bipartmodpenaltytermvolumes}
\end{equation}
where
\begin{eqnarray}
	\redlabeldegree{} & = & \vertsum{u} \reddegree{u} \samelabel{u}{} 
	\mathcomma \\
	\bluelabeldegree{} & = & \vertsum{u} \bluedegree{u} \samelabel{u}{} \label{eq:bluelabeldegreedef}
	\mathperiod
\end{eqnarray}
\Eqns{eq:bipartmodpenaltytermvolumes} \througheqn{eq:bluelabeldegreedef} adapt \eqns{eq:totaldegreeforlabel} \andeqn{eq:modpenaltytermlabelvolume} to bipartite networks.

We can rewrite \( \bppenaltyterm \) as
\begin{eqnarray}
	\bppenaltyterm & = & \half \vertsum{\vertlabel{}} \largeparens{ \vertsum{u} \reddegree{u} \samelabel{u}{} \vertsum{v} \bluedegree{v} \samelabel{v}{} } \nonumber \\
	& = & \half \vertsum{u} \vertsum{v} \reddegree{u} \bluedegree{v} \samelabel{u}{v}
	\mathperiod
	\label{eq:bppenaltyterm}
\end{eqnarray}
Writing the full objective function, we have
\begin{equation}
	\constrainedobjfunc = \half \vertsum{u} \vertsum{v} \largeparens{ \adjelem{u}{v} - \penaltywt \reddegree{u} \bluedegree{v} } \samelabel{u}{v}
	\mathperiod
	\label{eq:bpobjfunc}
\end{equation}
With 
\begin{equation}
	\penaltywt =  \frac{2}{\numedges}
	\mathcomma
\end{equation}
\eqn{eq:bpobjfunc} becomes
\begin{equation}
	\constrainedobjfunc = \numedges\bpmodularity
	\mathperiod
\end{equation}
The label propagation 
rule can again be used to maximize the bipartite modularity; we denote the algorithm as \LPAb. 
Implementation issues for \LPAb are treated in appendix~\sxnnum{sec:lpabimplementation}.

\section{Applications and Performance} \label{sec:modmax}

\subsection{Unipartite networks} \label{sec:unipartitemodmax}

We now turn to a comparison of the quality of solutions found by the various label
propagation algorithms discussed above. To quantify the solution quality, 
we will focus principally on the modularity \( \modularity \), although it is not strictly 
optimized except by \LPAm. Along with the \LPA, \LPAr,
and \LPAm variants discussed above, we will additionally consider
a hybrid algorithm, consisting of the standard \LPA followed by
optimization with \LPAm. The hybrid approach ensures that we are
at a maximum in the modularity, rather than just finding a
solution that hopefully offers a high value of \( \modularity \).

To begin, we apply the algorithms to randomly generated networks
with a known community structure. The most typical such class of
networks, introduced by \citet{GirNew:2002}, consists of four
communities, each containing 32 vertices. Edges exist between pairs
of vertices belonging to the same community with probability 
\( \problinkintra \) and between all other pairs of vertices with
probability \( \problinkinter \). The probabilities \( \problinkintra \) 
and \( \problinkinter \) are set so as to preserve the average
degree \( \meandegree \) of the vertices at a value of 16, while varying the average
number of edges~\( \meanoutdeg \) between a vertex and members of
other communities. As \( \meanoutdeg \) increases, the communities become 
increasingly difficult to identify. Although these model networks differ significantly from
real networks with community structure \citep{LanForRad:2008},
they do provide a simple initial test of community detection
algorithms.

\begin{figure}
	\includegraphics[width=\columnwidth]{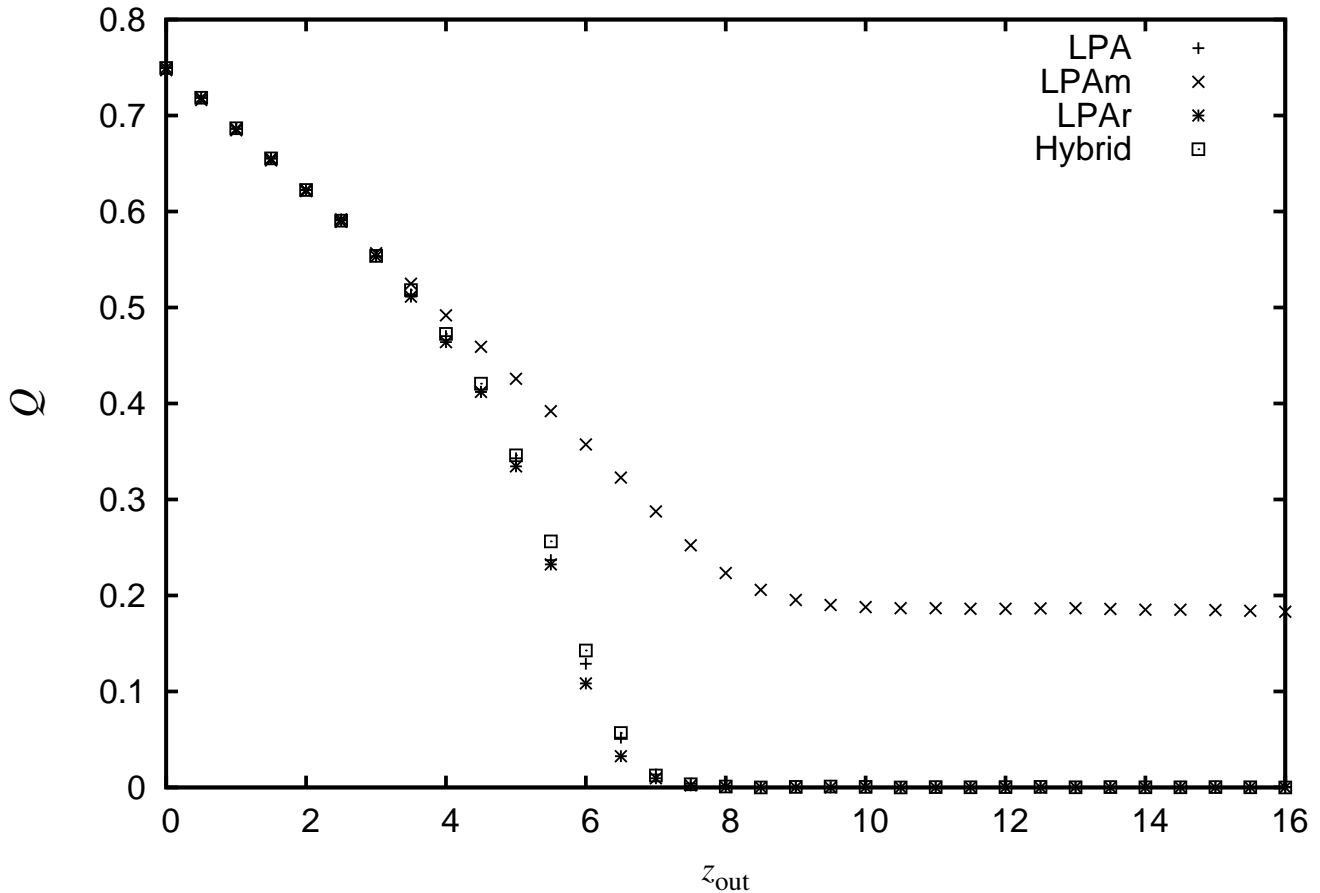}
	\caption{Modularity \( \modularity \) of community solutions from random 
	networks with known community structures. Each point shows the 
	average \( \modularity \) over 1000 instances of the random networks in 
	relation to the average number  
	\( \meanoutdeg \) of inter-community links for each vertex. 
	The hybrid algorithm consists of allowing the standard \LPA to run its course and
	find a solution, followed by application of \LPAm to the \LPA solution in order to ensure 
	that a local maximum of \( \modularity \) is reached.
	Error bars are smaller than the points. }
	\label{fig:blockmod}
\end{figure}

In \fig{fig:blockmod}, we show the modularity values for communities
found by the four algorithms. Each point shown gives the average
modularity from communities found in 1000 instances of the random
network model. As expected, \( \modularity \) drops as \( \meanoutdeg \) 
increases.

Since we know the actual communities for the model networks, we may
additionally assess the accuracy of the label assignments by directly
comparing to the known values.  We use the normalized mutual
information \( \normmutinfo \) \citep{DanDiaDucAre:2005} for the
comparison.  
Consider two schemes \(\schemex\) and \(\schemey\) for dividing the
\(\numvertices\) vertices into community groups. The probability
\(\probability{X=x, Y=y}\) that a vertex is assigned to community
\(x\) in scheme \(\schemex\) and to community \(y\) in scheme
\(\schemey\) is taken to be proportional to the size of the intersection between 
the sets of vertices
\( \commsetx \) and \( \commsety \) constituting the communities,
so that
\begin{equation}
	\probability{X=x, Y=y} = \frac{\largestraight{\commsetx \intersection \commsety}}{\numvertices}
	\label{eq:relateintersectiontoprobability}
	\mathperiod
\end{equation}
 Using the probability as defined in \eqn{eq:relateintersectiontoprobability}, we can calculate the normalized mutual information as
\begin{equation}
	\normalizedmutinfo{X}{Y} = \frac{2\mutinfo{X}{Y}}{\entropy{X} + \entropy{Y}}
	\label{eq:defnormmutinfo}
	\mathperiod
\end{equation}
\Eqn{eq:defnormmutinfo} is expressed in terms of the usual mutual information \(\mutinfo{X}{Y}\) and entropies \(\entropy{X}\) and  \(\entropy{X}\) \citep{CovTho:1991}, defined as
\begin{eqnarray} 
	\mutinfo{X}{Y} & = & \relentsum{\probability{X, Y}}{\probability{X}\probability{Y}}{x,y} \label{eq:defmutinfo}\\
	\entropy{X} & = & \entropysum{\probability{X}}{x} \label{eq:defentropyx}\\
	\entropy{Y} & = & \entropysum{\probability{Y}}{y} \label{eq:defentropyy}
	\mathperiod
\end{eqnarray} 
In \eqns{eq:defnormmutinfo} \througheqn{eq:defentropyy}, we have
made use of the common shorthand abbreviations \(\probability{X=x,
Y=y} = \probability{X, Y}\), \(\probability{X=x} = \probability{X}\),
and \(\probability{Y=y} = \probability{Y}\). The base of the
logarithms in \eqns{eq:defmutinfo} \througheqn{eq:defentropyy} is
arbitrary, as the computed measures only appear in the ratio in
\eqn{eq:defnormmutinfo}.

\begin{figure}
	\includegraphics[width=\columnwidth]{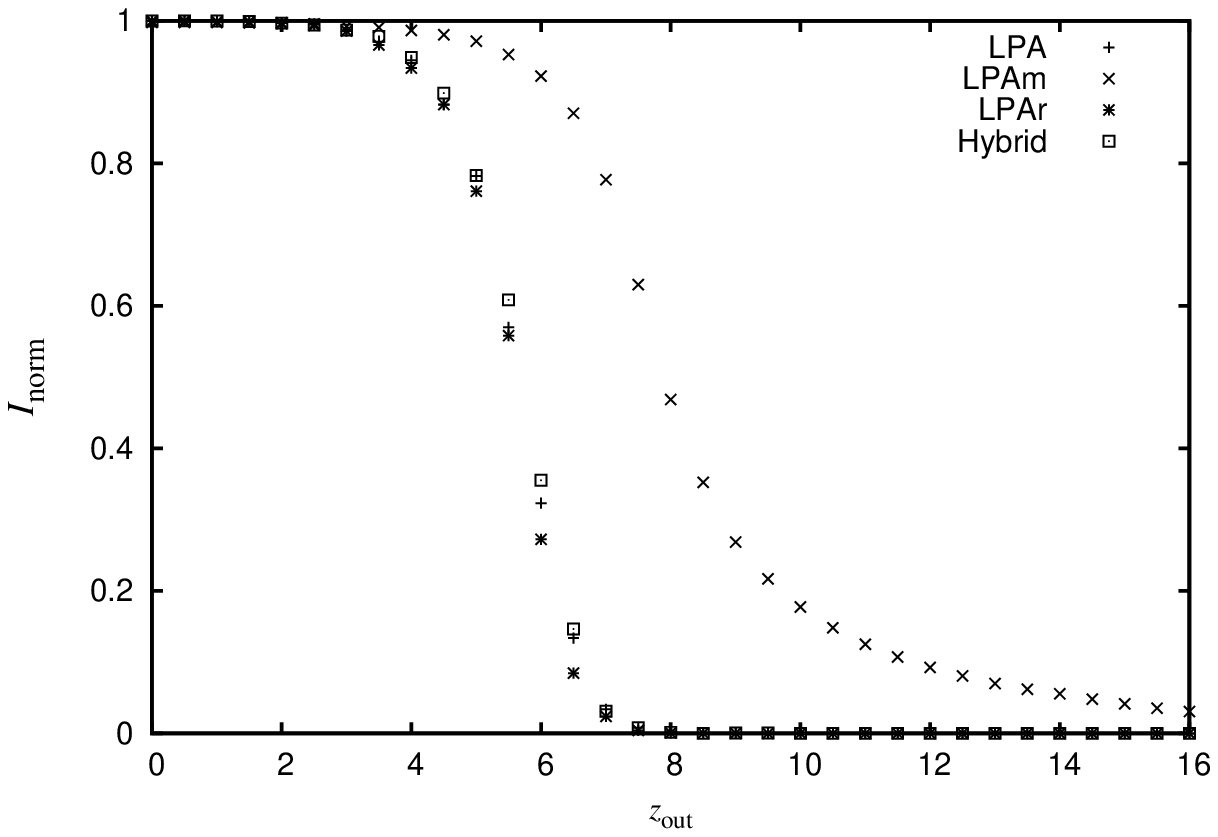}
	\caption{Accuracy of community solutions from random 
	networks with known community structures. Accuracy is
	quantified by the normalized mutual information \( \normmutinfo \) between the
	found and actual community solutions. Each point shows the
	normalized mutual information \( \normmutinfo \) over 1000 instances of the 
	random networks in relation to the average number  
	\( \meanoutdeg \) of inter-community links for each vertex. 
	The hybrid algorithm consists of allowing the standard \LPA to run its course and
	find a solution, followed by application of \LPAm to the \LPA solution in order to ensure 
	that a local maximum of \( \modularity \) is reached.
	Error bars are smaller than the points. }
	\label{fig:blockmutinfo}
\end{figure}

The normalized mutual information allows us to measure
the amount of information common to two different partitioning
schemes. Accordingly, we can explore the efficacy of the algorithm by taking 
one of the partitions to be the known modular
structure of the model networks and the other to be the structure
found using label propagation. When the found modules match the real ones, 
we have \(\normmutinfo=1\), and when they are
independent of the real ones, we have \(\normmutinfo=0\). Thus, as
\( \meanoutdeg \) increases, we expect \( \normmutinfo \) to decrease.
In \fig{fig:blockmutinfo}, we present values of \( \normmutinfo \)
from comparison of the real communities to the same community solutions used
for the \( \modularity \) calculations in \fig{fig:blockmod},
observing the expected decrease from \(\normmutinfo=1\) to
\(\normmutinfo=0\).

From \figs{fig:blockmod} \andfig{fig:blockmutinfo}, it is tempting
to conclude that \LPAm is superior to the other label propagation
variants. However, this conclusion is not borne out when the
algorithms are applied to real networks. In \tbl{tbl:networkdescription}, we 
list several networks that we have investigated
using the label propagation algorithms. 
The networks considered are a network of friendships between 
members of a university karate club \citep{Zac:1977}; 
a network of frequent associations between dolphins living near 
Doubtful Sound, New Zealand \citep{LusSchBoiHaaSloDaw:2003}; 
a network of collaborations between jazz musicians \citep{GleDan:2003}; 
a network of co-authorships for scientific papers concerning 
networks \citep{New:2006a}; 
and a network of co-authorships for scientific preprints posted to the condensed
matter archive \citep{condmat} between the years 1995 and 2003 \citep{New:2004a}. 
We give their sizes in terms of the number of vertices \( \numvertices \) and number of 
edges \( \numedges \). To indicate the degree to which the networks 
feature community structures, we also provide the modularity \( \modularity \), 
as determined using a greedy agglomerative hierarchical (\GAH)
method based on that of \citet{ClaNewMoo:2004}, wherein pairs 
of communities are successively merged so as to cause the 
largest possible increase in \( \modularity \) at each step.
While edge weights are available in some 
cases, in this work we uniformly treat all network edges 
as unweighted.

\begin{table}
	\begin{ruledtabular}
	\begin{tabular}{lrrr}
		\multicolumn{1}{l}{Network} & \multicolumn{1}{r}{\(\numvertices\)} & \multicolumn{1}{r}{\(\numedges\)} & \( \modularity \)\\ 
		\hline
		karate & 34 & 78 & 0.3807\\ 
		dolphins & 62 & 159 & 0.4923 \\ 
		jazz & 198 & 2742 & 0.4389 \\ 
		network science & 1589 & 2742 & 0.9555 \\
		condmat 2003 & 31163 & 120029 & 0.6885 \\ 
	\end{tabular}
	\end{ruledtabular}
	\caption{Basic properties of networks used to test label propagation algorithm variants. 
	The sizes of the network are described by the number of vertices \( \numvertices \) and 
	number of edges \( \numedges \). Each network has significant modular character, as 
	indicated by the modularity \( \modularity \).
	}
	\label{tbl:networkdescription}
\end{table}

For each of the networks, we identify communities using each of the algorithms \LPA, \LPAm, and \LPAr. Additionally, we consider a hybrid algorithm consisting of \LPA 
followed by \LPAm, thus ensuring that we are at a maximum of the \( \modularity \).
We applied each of the four algorithms one hundred times to each of the networks. In 
\tbl{tbl:combinedmaxperfcomp}, we show the maximum modularity found in the samples, 
suggesting the potential performance, while in \tbl{tbl:combinedmeanperfcomp},
we show the mean modularity, revealing the expected performance. From the tables, we 
can see that no algorithm variant is clearly superior, suggesting that the four variants all
explore slightly different portions of the solution space. Interestingly, the \LPAr variant,
which worked poorly when applied to the Southern women network (\sxn{sec:drawbacks}),
provides the best results on the two large co-authorship networks.
We note that the label propagation variants produce community solutions with
modularity values similar to those found with the \GAH approach and shown 
in \tbl{tbl:networkdescription}.

\begin{table}
	\begin{ruledtabular}
	\begin{tabular}{lllll}
		\multicolumn{1}{l}{Network} & \multicolumn{1}{c}{\LPA} & \multicolumn{1}{c}{\LPAm} & \multicolumn{1}{c}{\LPAr} & \multicolumn{1}{c}{Hybrid} \\ 
		\hline
		karate & 0.4156 & 0.4000 & 0.4156  &  0.4198 \\ 
		dolphins & 0.5237 & 0.5157 & 0.5265 & 0.5253 \\
		jazz & 0.4424 & 0.4448 & 0.4428 & 0.4442 \\
		network science & 0.8924 & 0.8723 & 0.9163 & 0.8934 \\
		condmat 2003 & 0.6228 & 0.5947 & 0.6578 & 0.6360 \\ 
	\end{tabular}
	\end{ruledtabular}
	\caption{Maximum modularity \( \modularity \) found for network community assignments. 
	Values were calculated
	using one hundred samples for each network for each of the standard \LPA, \LPAm,
	\LPAr, and a hybrid approach consisting of maximization with \LPA
	followed by maximization with \LPAm.}
	\label{tbl:combinedmaxperfcomp}
\end{table}

\begin{table}
	\begin{ruledtabular}
	\begin{tabular}{lllll}
		\multicolumn{1}{l}{Network} & \multicolumn{1}{c}{\LPA} & \multicolumn{1}{c}{\LPAm} & \multicolumn{1}{c}{\LPAr} & \multicolumn{1}{c}{Hybrid} \\ 
		\hline
		karate & 0.366(6) & 0.347(3) & 0.352(9) & 0.386(4) \\ 
		dolphins & 0.484(4) & 0.4956(8) & 0.484(5) & 0.495(3) \\
		jazz & 0.336(9) & 0.4351(9) & 0.34(1) & 0.366(7) \\
		network science & 0.8792(6) & 0.8618(5) & 0.9046(5) & 0.8806(6) \\
		condmat 2003 & 0.6073(6) & 0.5828(4) & 0.6420(6) & 0.6139(9) \\ 		
	\end{tabular}
	\end{ruledtabular}
	\caption{Mean modularity \( \modularity \) found for network community assignments. 
	Values were calculated
	using one hundred samples for each network for each of the standard \LPA, \LPAm,
	\LPAr, and a hybrid approach consisting of maximization with \LPA
	followed by maximization with \LPAm. The uncertainty of the final digit, calculated as the 
	standard error of the mean, is shown parenthetically.}
	\label{tbl:combinedmeanperfcomp}
\end{table}

\subsection{Bipartite networks}

As we did above for unipartite networks, we next quantify the quality
of community solutions found in bipartite networks. We measure
community quality using the bipartite modularity \( \bpmodularity \), 
calculating values for the \LPA, \LPAr, and \LPAb variants.
Again, we consider a hybrid algorithm, consisting of \LPA followed
by \LPAb, ensuring that the solutions are at maxima in \( \bpmodularity \).

We examine the performance using four real-world bipartite networks.
The networks are the Southern women network, described above in
\sxn{sec:drawbacks}; a network describing corporate interlocks in
Scotland, based on the membership of boards of directors for Scottish
firms during 1904--5 \citep{ScoHug:1980}; and bipartite versions
of the condensed matter and network science co-authorship networks
considered in \sxn{sec:unipartitemodmax}, including authors and
their papers as the two parts of the network. 
In \tbl{tbl:bipartitenetworkdescription}, we indicate the size and extent of 
community structure in the networks.
We show the size using the number of 
vertices~\( \numredvertices \) and~\( \numbluevertices \) in the 
two parts of the networks, as well as the number of 
edges~\( \numedges \). We show the extent of community structure 
using the bipartite modularity \( \bpmodularity \), as determined using a 
greedy agglomerative hierarchical method, analogous to that commonly used for 
unipartite networks \citep{New:2004a,ClaNewMoo:2004}.

\begin{table}
	\begin{ruledtabular}
	\begin{tabular}{lrrrr}
		\multicolumn{1}{l}{Network} & \multicolumn{1}{r}{\(\numredvertices\)} & \multicolumn{1}{r}{\(\numbluevertices\)} & \multicolumn{1}{r}{\(\numedges\)} & \( \bpmodularity \)\\ 
		\hline
		Southern women & 14 & 18 & 89 & 0.3430 \\
		Scotland interlocks  & 108 & 136 & 358 & 0.6969 \\
		network science & 959 & 1588 & 2580 & 0.9695 \\
		condmat 2003 & 31162 & 47055 & 134600 & 0.8700 \\ 
	\end{tabular}
	\end{ruledtabular}
	\caption{Basic properties of bipartite networks used to test label propagation 
	algorithm variants. The sizes of the network are described by the numbers of 
	vertices \( \numredvertices \) and \( \numbluevertices \) in the two parts of the network
	and by the
	number of edges \( \numedges \). Each network has significant modular character, as 
	indicated by the bipartite modularity \( \bpmodularity \).}
	\label{tbl:bipartitenetworkdescription}
\end{table}

To each network, we apply each label propagation algorithm one hundred times. The
maximum and mean values found for \( \bpmodularity \) are given in
\tbls{tbl:combinedbpmaxperfcomp} \andtbl{tbl:combinedbpmeanperfcomp},
respectively. For the Southern women network, we note that \LPAr
is clearly the worst of the algorithms considered, consistent with
its tendency to assign the same label to all vertices, as seen in
\fig{fig:optimizeLPA}. Further, the improved performance of \LPAb
on the Southern women network in terms of the average \( \bpmodularity \) 
indicates that the inclusion of \(
\bppenaltyterm \) reduces the frequent appearance of the trivial
solution with all vertices in the same community.

\begin{table}
	\begin{ruledtabular}
	\begin{tabular}{lllll}
		\multicolumn{1}{l}{Network} & \multicolumn{1}{c}{\LPA} & \multicolumn{1}{c}{\LPAb} & \multicolumn{1}{c}{\LPAr} & \multicolumn{1}{c}{Hybrid} \\ 
		\hline
		Southern women & 0.3212 &	0.3192  & 0.3184 & 0.3257 \\
		Scotland interlocks & 0.5782 & 0.5783 & 0.6552 & 0.5975 \\
		network science & 0.8137 & 0.7807 & 0.8948 & 0.8172 \\
		condmat 2003 & 0.6378 & 0.6179 & 0.7232 & 0.6587 \\
	\end{tabular}
	\end{ruledtabular}
	\caption{Maximum bipartite modularity \( \bpmodularity \) found for bipartite network community assignments. 
	Values were calculated
	using one hundred samples for each network for each of the standard \LPA, \LPAb,
	\LPAr, and a hybrid approach consisting of maximization with \LPA
	followed by maximization with \LPAb.
	}
	\label{tbl:combinedbpmaxperfcomp}
\end{table}

\begin{table}
	\begin{ruledtabular}
	\begin{tabular}{lllll}
		\multicolumn{1}{l}{Network} & \multicolumn{1}{c}{\LPA} & \multicolumn{1}{c}{\LPAb} & \multicolumn{1}{c}{\LPAr} & \multicolumn{1}{c}{Hybrid} \\ 
		\hline
		Southern women & 0.19(1) & 0.250(3) & 0.17(1) & 0.27(1) \\
		Scotland interlocks & 0.543(1) & 0.548(2) & 0.633(1) & 0.568(1) \\
		network science &  0.788(1) & 0.7624(6) & 0.8733(8) & 0.7986(8) \\
		condmat 2003 & 0.6314(3) & 0.6142(1) & 0.7183(2) & 0.6536(2) \\
	\end{tabular}
	\end{ruledtabular}
	\caption{Mean bipartite modularity \( \bpmodularity \) found for bipartite network community assignments. 
	Values were calculated
	using one hundred samples for each network for each of the standard \LPA, \LPAb,
	\LPAr, and a hybrid approach consisting of maximization with \LPA
	followed by maximization with \LPAb. 
	}
	\label{tbl:combinedbpmeanperfcomp}
\end{table}

Despite the success of \LPAb on the Southern women network, it is
less successful on the other networks. Performance is quite similar
for \LPA and \LPAb on the Scotland corporate interlocks network,
but \LPAb is otherwise outperformed by the other label propagation
variants. Indeed, \LPAr provides the best results for the larger
networks, in contrast to its poor results for the Southern women
network.
Values of \( \bpmodularity \) for community solutions found using
the label propagation variants are generally somewhat less than the
values, shown in \tbl{tbl:bipartitenetworkdescription}, for communities
found using a greedy agglomerative hierarchical approach. 

\section{Discussion} \label{sec:discussion}

We have examined the label-propagation algorithm as an optimization
problem, identifying  community solutions that it finds with the
maxima of an objective function. The objective function, which is
just the number of network edges connecting vertices with the same
labels, has the significant conceptual drawback that increasing the
objective function need not produce what we would consider to be
better communities. Markedly, the globally optimal solution is
completely uninformative, with all vertices in the same community.
Label propagation thus depends on reaching one of the large number
of local maxima in the objective function to avoid the trivial
global solution. Attempts to improve on the algorithm
may be counterproductive, giving less information while reaching
nominally better solutions. By modifying the objective function,
we defined several label-propagation algorithms that are constrained
to avoid assigning all vertices to the same community. One of the
constrained label-propagation algorithms, \LPAm, finds local maxima in the 
modularity \( \modularity \); another, \LPAb, finds local maxima in a 
modified modularity \( \bpmodularity \)
for bipartite networks.

Although formally equivalent, there are important conceptual
differences between the usual definition of the modularity \(
\modularity \) in terms of a null model network and the version
based on constraints presented here. For example, the parameter \(
\penaltywt \) seems quite arbitrarily chosen in the constraint-based
version. In fact, the community solutions found by \LPAm are not especially 
sensitive to the choice of \( \penaltywt \).  The value can, for instance, be 
cut in half to \( \penaltywt = 1/4\numedges \) with significant change only in the case of
the mean modularity for Zachary's karate network---in which 
the mean modularity value actually increases by about 10\%. 

More significantly, the constraint as given in
\eqn{eq:modpenaltytermlabelvolume} makes clear that modularity
favors communities of similar size, with size measured by the total
degree of the vertices in the community. As the distribution of
community sizes may be far from uniform (see, for example, Fig.~3
in Ref.~\cite{ClaNewMoo:2004}), the constraint approach points
immediately towards a practical difficulty in detecting community
by maximizing modularity. In contrast, difficulties due to varying 
community sizes were recognized \citep{DanDiaAre:2006} only some 
time after the original introduction of modularity using a null model.

Corresponding properties hold in the case of the bipartite modularity
\( \bpmodularity  \). Again, \( \penaltywt \) seems arbitrarily
chosen; halving the parameter value to \( \penaltywt = 1/\numedges\) again 
only causes a significant change for the small Southern women network, 
increasing the mean bipartite modularity found by about 10\%. 
In the bipartite case, communities of similar size are also
favored, but the relevant size is now the geometric mean of the
total degrees within the community for the two parts of the network,
as seen in \eqn{eq:bipartmodpenaltytermvolumes}. We thus expect
that community identification methods based on maximizing  
\( \bpmodularity  \) will also have difficulties with networks consisting
of communities of diverse sizes. Although this latter fact has been
anticipated \citep{Bar:2007} based on parallels to the unipartite
case, it has not been previously demonstrated.

In light of the results for the real-world networks
(\tbls{tbl:combinedmaxperfcomp} \andtbl{tbl:combinedmeanperfcomp} for unipartite networks, 
\tbls{tbl:combinedbpmaxperfcomp} \andtbl{tbl:combinedbpmeanperfcomp} 
for bipartite networks),
it seems clear that the main label propagation variants we
have considered---\LPA, \LPAm, \LPAr, \LPAb---all give good community
results. The performance differences indicate that the algorithm
variants explore slightly different portions of the community
solution space.  No variant is clearly superior, which is not
surprising given that we are trying to identify communities without
prior information on their number, size, or nature. 

When compared to the modularity values for community solutions
generated by greedy agglomerative hierarchical methods, the label
propagation variants appear to provide no advantage or, in the case
of bipartite networks, to entail a distinct disadvantage.  We stress that the
difference in modularity values should not be overvalued, for two
main reasons. First, the modularity measure, while popular, is not
the only possibility, nor is it without drawbacks (see, \eg,
Ref.~\cite{ForBar:2007}). Second, the algorithms are quite different,
so no single point of comparison will be determinative in general.
A more thorough characterization of performance is needed to 
establish reliable guidelines for choosing appropriate algorithms to analyze
particular networks; this will be the subject of future work.

The performance of \LPAm is especially interesting: although it is the
only variant directly maximizing \( \modularity \), other variants
produce better results in terms of \( \modularity \) for some of the
networks considered.  This appears to be due to a fundamental difference in
the role played by the modularity in the algorithm variants.  Lacking
an objective function, \citet{RagAlbKum:2007} used the modularity
of the final community solution to assess the acceptability of their
\LPA, as did we when assessing \LPAr in the present work. Thus, in
\LPA and \LPAr, the modularity is used diagnostically to select a
best result from candidate solutions produced based on other
considerations.  In contrast, the modularity plays an essential
role in \LPAm, impacting the final community solution as well as the
intermediate community states reached during the course of the algorithm.
The dynamical path followed through the space of label assignments
is driven to favor states where all communities are similar in
total degree, although there is little reason to believe such paths are universally
ideal or particularly free of local maxima.  
Thus, the null model network used in defining the
modularity---regardless of its suitability as a model of the final
communities---may be an impractical model of the intermediate
communities. This might be addressed by varying \( \penaltyterm \),
gradually introducing the penalty term \( \modpenaltyterm \) and
thus the null network mode. Similar considerations hold for \LPAb
and the corresponding null network models for bipartite networks.

Overall, we have found the label propagation algorithm to be a 
promising approach to understanding networks, with a number of
desirable qualities.  Label propagation seems well suited as a basis
for more specialized community-detection methods, as well as application to other aspects
of networks besides community structure.  A clear understanding of
the drawbacks of label propagation, as well as its strengths, will help to avoid problems
and facilitate further applications.

\acknowledgments

We thank Mark Newman for providing the bipartite versions of networks 
describing co-authorships in condensed matter and in network sciences.
This work has been supported by the European FP6-NEST-Adventure Programme, 
contract number 028875.


\bibliographystyle{apsrev} 

\appendix
\section{A label-propagation algorithm for maximizing modularity} 
\label{sec:lpamimplementation}

The label propagation algorithm presented by \citet{RagAlbKum:2007}
has desirable performance properties. Each relabeling iteration
through the vertices has a computational (time) complexity 
\(\order{\numedges} \) linear in the number of edges \( \numedges \)
in the graph.  For many networks, the number of vertices 
\(\numvertices \) scales with the number of edges, so the computational
complexity for each relabeling step can instead be given as 
\(\order{\numvertices} \).

As seen in \sxn{sec:constrainedlpa}, the objective function for the
\LPA can be constrained to reproduce the modularity. Consequently,
it is necessary to adapt the algorithm itself to obtain an efficient
procedure for maximizing the modularity. Modifications can be made
so as to maintain the \( \order{\numedges} \) time complexity. Here, we 
consider the constraint \( \modpenaltyterm \) given in \eqn{eq:modpenaltyterm}, \ie, we implement \LPAm.

First, consider the objective function from \eqn{eq:objfunction}.
Recall that the \LPA update rule (\eqn{eq:lparuleadj}) can be applied
with any symmetric matrix \( \arbsymelem{u}{v} \) playing the role
of the adjacency matrix \( \adjelem{u}{v} \) (see \sxn{sec:objfunc}).
Further, it is clear that the objective function may be shifted by adding an
arbitrary constant \( \arbconst \) without altering the locations
of the maxima in the space of label assignments. By setting 
\(\arbconst = -\vertsum{u} \arbsymelem{u}{u} \), we eliminate the
diagonal elements \( \arbsymelem{u}{u} \) from consideration,
producing an objective function
\begin{equation}
	\objfunc = \vertsum{v} \excvertsum{u}{v} \arbsymelem{u}{v} \samelabel{u}{v}
\end{equation}
and update rule
\begin{equation}
	\newvertlabel{v} = \argmax{\vertlabel{}} \excvertsum{u}{v} \arbsymelem{u}{v} \samelabel{u}{}
	\mathperiod
\end{equation}
The above transformation eliminates constant self-interaction terms.

Next, identify \( \arbsymelem{u}{v} \) as 
\( \adjelem{u}{v} - \penaltywt \degree{u} \degree{v}  \) to match the \LPAm variant, giving
\begin{equation}
	\newvertlabel{v} = \argmax{\vertlabel{}} \excvertsum{u}{v} \largeparens{\adjelem{u}{v} - \penaltywt \degree{u} \degree{v}} \samelabel{u}{} 
\end{equation}
or, equivalently, 
\begin{equation}
	\newvertlabel{v} = \argmax{\vertlabel{}}  \largeparens{\excvertsum{u}{v} \adjelem{u}{v} \samelabel{u}{} - \penaltywt \degree{v} \excvertsum{u}{v} \degree{u} \samelabel{u}{}}
	\label{eq:lpamodrulebasis}
	\mathperiod
\end{equation}
The first sum in \eqn{eq:lpamodrulebasis} corresponds 
to the counting of labels on neighboring vertices in the original 
label propagation algorithm. Write this as
\begin{equation}
	\numnbrswithlabel{v}{} = \excvertsum{u}{v} \adjelem{u}{v} \samelabel{u}{}
	\mathperiod
	\label{eq:countneighborlabels}
\end{equation}
The second sum in \eqn{eq:lpamodrulebasis} can be rewritten as
\begin{equation}
	 \excvertsum{u}{v} \degree{u} \samelabel{u}{} = \labeldegree{} - \degree{v} \samelabel{v}{}
	 \mathcomma
\end{equation}
where
\begin{equation}
	\labeldegree{} = \vertsum{u} \degree{u} \samelabel{u}{}
	\label{eq:labelvol}
	\mathperiod
\end{equation}
Analogously to the volume of a graph, \( \labeldegree{}  \) can 
be viewed as a sort of volume for the labels. 

Incorporating \eqns{eq:countneighborlabels} \andeqn{eq:labelvol} into \eqn{eq:lpamodrulebasis}, we obtain
\begin{equation}
	\newvertlabel{v} = \argmax{\vertlabel{}}  \largeparens{\numnbrswithlabel{v}{} - \penaltywt \degree{v} \labeldegree{} + \penaltywt \degree{v}^2 \samelabel{v}{}}
	\label{eq:lpamodrule}
	\mathperiod
\end{equation}
The modified label propagation rule, as expressed in 
\eqn{eq:lpamodrule}, can be readily implemented so that each 
pass through the vertices requires  \( \order{\numedges} \) 
worst-case time complexity. 

The algorithm is initialized by assigning a unique numerical label
\( \vertlabel{} \) to each vertex and by setting \( \labeldegree{} \) 
to the degree of the vertex. The first term, \( \numnbrswithlabel{v}{} \), 
requires that the labels of the neighbors for each vertex be
counted and is thus \( \order{\numedges} \); this is unsurprising
as it is equivalent to the unmodified label propagation algorithm,
which is \( \order{\numedges} \). The second term appears to require
that each possible label be checked for each vertex, giving 
\( \order{\numvertices^2} \). However,  it is only necessary to consider
the labels of the neighbors for each vertex---no other label can
make a positive contribution to the modularity, but a zero contribution
can be had by assigning an unused label. A list of unused labels
can be kept, allowing \( \order{1} \)  access. Additionally, the
\( \labeldegree{} \) must be updated if the label changes,
but this is also \( \order{1} \) for each vertex. In total, checking and 
updating the \( \labeldegree{} \) terms for all vertices is 
\( \order{\numedges} \). The final term in \eqn{eq:lpamodrule} is
\( \order{\numvertices} \) in total. With all three terms taken
into account, the modified algorithm thus has worst-case 
\( \order{\numedges} \) time complexity.

\section{A label-propagation algorithm for maximizing bipartite modularity} 
\label{sec:lpabimplementation}

In \eqn{eq:bpobjfunc}, we have presented an objective function corresponding to the bipartite modularity \( \bpmodularity \), with form
\begin{equation}
	\constrainedobjfunc = \half \vertsum{u} \vertsum{v} \largeparens{ \adjelem{u}{v} - \penaltywt \reddegree{u} \bluedegree{v} } \samelabel{u}{v}
	\mathperiod
	\label{eq:bpobjfuncrepeat}
\end{equation}
We cannot directly apply the label propagation update rule from
\eqn{eq:lparuleadj}, as \( \adjelem{u}{v} - \penaltywt \reddegree{u}
\bluedegree{v} \) is in general asymmetric. Despite this, we can
define a label propagation rule for \( \constrainedobjfunc \). 

We
rewrite \eqn{eq:bpobjfunc} by first taking advantage of the symmetry
of \( \adjelem{u}{v} \) and \( \samelabel{u}{v} \), giving
\begin{equation}
	\constrainedobjfunc = \half \vertsum{u} \vertsum{v} \largeparens{ \adjelem{v}{u} - \penaltywt \reddegree{u} \bluedegree{v} } \samelabel{v}{u}
	\mathperiod
\end{equation}
Next, we switch the dummy indices \( u \) and \( v \), resulting in
\begin{equation}
	\constrainedobjfunc = \half \vertsum{u} \vertsum{v} \largeparens{ \adjelem{u}{v} - \penaltywt \reddegree{v} \bluedegree{u} } \samelabel{u}{v}
	\mathperiod
	\label{eq:bpobjfuncswapped}
\end{equation}
Averaging \eqns{eq:bpobjfuncrepeat} \andeqn{eq:bpobjfuncswapped}, we obtain
\begin{equation}
	\constrainedobjfunc = \half \vertsum{u} \vertsum{v} \largeparens{ \adjelem{u}{v} - \frac{\penaltywt}{2} \largeparens{\reddegree{u} \bluedegree{v}  + \reddegree{v} \bluedegree{u} }} \samelabel{u}{v}
	\mathcomma
	\label{eq:bpobjfuncsym}
\end{equation}
which is in terms of a symmetric matrix and thus suitable for use with \eqn{eq:lparuleadj}.

The objective function, as expressed in \eqn{eq:bpobjfuncsym}, can
be converted into the \LPAb label propagation rule for bipartite
modularity in a fashion directly parallel to that presented in
appendix~\sxnnum{sec:lpamimplementation}. The resulting update rule
has the form
\begin{equation}
	\newvertlabel{v} = \argmax{\vertlabel{}}  \largeparens{\numnbrswithlabel{v}{} 
		- \frac{\penaltywt \bluedegree{v}}{2} \redlabeldegree{}  
		- \frac{\penaltywt \reddegree{v}}{2} \bluelabeldegree{} 
		+ \frac{\penaltywt}{2} \reddegree{v}^2 \samelabel{v}{} 
		+ \frac{\penaltywt}{2} \bluedegree{v}^2 \samelabel{v}{}}
	\label{eq:lpabpmodrule}
	\mathcomma
\end{equation}
where
\begin{eqnarray}
	\redlabeldegree{} & = & \vertsum{u} \reddegree{u} \samelabel{u}{} 
	\mathcomma \\
	\bluelabeldegree{} & = & \vertsum{u} \bluedegree{u} \samelabel{u}{} 
	\mathperiod
\end{eqnarray}
By updating \( \redlabeldegree{} \) and \( \bluelabeldegree{} \) when labels change, the algorithm can be implemented efficiently. The details, omitted here, are similar to those given in appendix~\sxnnum{sec:lpamimplementation} and result in the same \( \order{\numedges} \) worst-case time complexity for each iteration of \LPAb.

\end{document}